\documentclass[runningheads]{llncs}
\usepackage[T1]{fontenc}
\usepackage{amssymb}
\usepackage{amsmath}
\usepackage{multirow}
\usepackage{booktabs}
\usepackage{footnote}
\usepackage{marvosym}
\usepackage[colorlinks,urlcolor=blue,citecolor=blue,linkcolor=red]{hyperref}
\usepackage{graphicx,verbatim}
\usepackage[table]{xcolor}
\usepackage{color}
\definecolor{dino}{RGB}{249,231,227}

\begin{document}
\title{MurreNet: Modeling Holistic Multimodal Interactions Between Histopathology and Genomic Profiles for Survival Prediction}
\titlerunning{Modeling Holistic Multimodal Interactions for Survival Prediction}
\author{
Mingxin Liu\inst{1} \and 
Chengfei Cai\inst{2} \and 
Jun Li\inst{1} \and 
Pengbo Xu\inst{3} \and 
Jinze Li\inst{1} \and \\ 
Jiquan Ma\inst{4} \and 
Jun Xu\inst{1}\textsuperscript{\(\left(\textrm{\Letter}\right)\)} 
}
\authorrunning{M. Liu et al.}
\institute{
Jiangsu Key Laboratory of Intelligent Medical Image Computing, \\ School of Artificial Intelligence, \\ Nanjing University of Information Science and Technology, Nanjing, China \email{jxu@nuist.edu.cn} \and
College of Information Engineering, Taizhou University, Taizhou, China \and
College of Bioinformatics Science and Technology, \\ Harbin Medical University, Harbin, China \and
School of Computer and Big Data, Heilongjiang University, Harbin, China \\
}
    
\maketitle              
\begin{abstract}
Cancer survival prediction requires integrating pathological Whole Slide Images (WSIs) and genomic profiles, a challenging task due to the inherent heterogeneity and the complexity of modeling both inter- and intra-modality interactions. Current methods often employ straightforward fusion strategies for multimodal feature integration, failing to comprehensively capture modality-specific and modality-common interactions, resulting in a limited understanding of multimodal correlations and suboptimal predictive performance. To mitigate these limitations, this paper presents a \textbf{Mu}ltimodal \textbf{R}ep\textbf{r}esentation D\textbf{e}coupling \textbf{Net}work (\textbf{MurreNet}) to advance cancer survival analysis. Specifically, we first propose a Multimodal Representation Decomposition (MRD) module to explicitly decompose paired input data into modality-specific and modality-shared representations, thereby reducing redundancy between modalities.
Furthermore, the disentangled representations are further refined then updated through a novel training regularization strategy that imposes constraints on distributional similarity, difference, and representativeness of modality features. Finally, the augmented multimodal features are integrated into a joint representation via proposed Deep Holistic Orthogonal Fusion (DHOF) strategy. 
Extensive experiments conducted on six TCGA cancer cohorts demonstrate that our MurreNet achieves state-of-the-art (SOTA) performance in survival prediction.
\keywords{Whole slide images \and Survival outcome prediction \and Multimodal learning \and Multimodal representation decoupling.}
\end{abstract}

\section{Introduction}
Survival analysis, a cornerstone of clinical practice, aims to predict the relative risk of death in cancer prognosis and provide crucial references for the cancer prevention and treatment plan~\cite{song2023artificial}. 
In current clinical practice, survival prediction methods integrating gigapixel WSIs and genomic data are particularly promising~\cite{songmultimodal}. 
WSIs provide qualitative morphological information and detailed spatial depiction of tumor tissues~\cite{liu2024exploiting,liu2024unleashing}, and their interactions in the tumor micro-environment (TME); whereas genomics offer molecular information from the microcosmic perspective. 
Both two modalities have their unique strengths and constraints, and their integration is pivotal for advancing the precision of survival analysis.
Therefore, the recent focus of survival analysis studies has shifted from single-modal predictions~\cite{cai2024pathologist,ching2018cox,klambauer2017self} to utilizing multimodal information~\cite{chen2021multimodal,chen2022pan,jaume2024modeling,liu2023mgct,xiong2024mome} since their superior capabilities in patient survival prognosis~\cite{cai2024seqfrt,wang2025histo}. 

The multimodal knowledge within integrative pathology-genomic data can be decomposed into
distinct components: modality-common (inter-modal) and modality-specific (intra-modal) representations.
Furthermore, these two divers features are extremely intricate, as each modality embodies wealthy information, yet only a small subset can be mutually correlated with cancer survival.
Therefore, previous studies focused on integrating common information, highlighting the alignment of shared features across modalities~\cite{chen2021multimodal,liu2023mgct,xiong2024mome,xu2023multimodal,zhou2023cross}.
For example, MGCT~\cite{liu2023mgct} uses mutual-guided cross-modality attention to model shared information and enhance modality-common interactions. 
While modality-common information often dominates multimodal fusion, resulting in the suppression of modality-specific information, thus neglecting the richness of distinct perspectives.
Recently, several works~\cite{long2024mugi,zhang2024prototypical,zhou2024cohort} have attempted to improve multimodal interactions modeling by a \textit{divide-and-conquer} manner, handling modality-specific and -common representations while preserving their unique contributions.
For instance, CCL~\cite{zhou2024cohort} presents a Cohort-individual Cooperative Learning framework for survival analysis by collaborating knowledge decomposition and cohort guidance.
MuGI~\cite{long2024mugi} leveraged a generator and a MLP to extract the shared and specific information respectively, then parallel cross-attentions to integrate the resulting features.
Nevertheless, these methods still suffer from some limitations: simply decoupling the features into shared and specific parts then simple fusion, lacking further optimization on multimodal fusion or dedicated learning strategies for guiding highly divergent features through targeted constraints, which resulting in a suboptimal performance.

\begin{figure}[h]
    \centering
\includegraphics[width=1.0\textwidth]{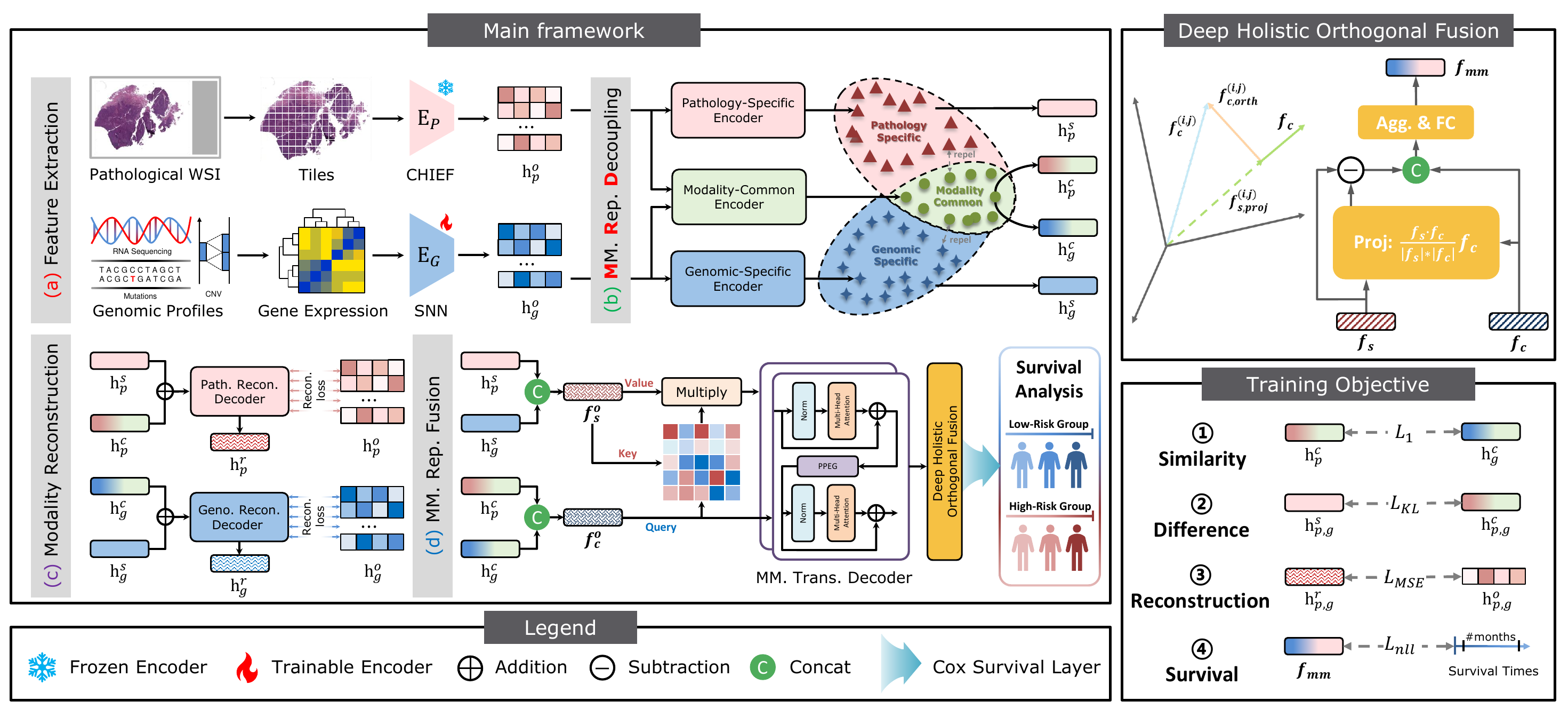}
\caption{Overview of the proposed MurreNet: (a) feature extraction for pathology and genomic features, (b) multimodal representation decoupling module, (c) modality reconstruction module, (d) multimodal representation fusion by deep holistic orthogonal fusion module and comprehensive training regularization strategy for cancer prognosis.} \label{framework}
\end{figure}

To tackle the challenges above, we propose the multimodal representation decoupling network (MurreNet), as depicted in \figurename~\ref{framework}, tailored for survival prediction by modeling holistic interactions between pathology and genomic data, which highlighting the importance of decomposed multimodal representation learning. 
Specifically, our key contributions are in three folds: 
(1) a Multimodal Representation Decomposition (MRD) module is devised to learn modality-specific and modality-common representations, providing a comprehensive and disentangled view of multimodal data.
(2) a Deep Holistic Orthogonal Fusion (DHOF) module is proposed to integrate complementary modality-specific and modality-common information.
(3) a robust loss combination of modality similarity, difference, reconstruction is introduced to provide comprehensive learning constraints.
Extensive experimental results confirmed that our proposed model outperforms other cutting-edge cancer survival prediction methods significantly.

\section{Methodology}
\subsection{Problem Formulation and Feature Extraction}
\subsubsection{Problem Formulation.}
In this multimodal survival analysis task, our objective is to learn a robust multimodal representation utilizing paired pathology $P$ and genomic data $G$, and develop a survival prediction model to estimate the hazard function $f_{hazard}(t)=f_{hazard}(T=t\vert T\geq t, (P,G)) \in [0,1]$, which represents the probability of death for a patient right after the discrete time point $t$ rather than estimating the survival time directly. Then, an ordinal value can be obtained via integrating the negated hazard functions:$f_{surv}(t|(P,G))=\prod_{u=1}^t(1-f_{hazard}(u))$.

\subsubsection{Pathological Feature Embedding.}
Due to the gigapixel nature of WSIs (e.g., 200,000$\times$100,000 pixels), we adopt the CLAM~\cite{lu2021data} to automatically partition each WSI into a series of non-overlapping $256\times256$ patches at 20$\times$ magnification. Subsequently, the pathological foundation model, CHIEF~\cite{wang2024pathology}, is employed in a frozen state to embed each patch into an instance-level feature vector. Finally, these patch-level features are converted into a 768-dimensional feature embedding $\mathrm{h}_{p}^{o}$ through a learnable fully-connected layer. 

\subsubsection{Genomic Feature Embedding.}
The genomic profiles consists of various 1$\times$1 values, including gene mutation status, RNA Sequencing, DNA methylation, $etc$. However, the data exhibits a high-dimensional low-sample size characteristic. To mitigate this issue, we follow~\cite{chen2021multimodal,xu2023multimodal} to group these high signal-to-noise ratio data into six genomic sequences (see~\ref{sec:datasets}). These grouped sequences are stacked then fed into a fully connected layer to obtain $d$-dimensional genomic embedding $\mathrm{h}_{g}^{o}$. 

\subsection{Multimodal Representation Decoupling}
To preform a comprehensive and disentangled perspective of multimodal representations, we develop three encoders within our MRD module. These encoders are designed to capture distinct types of knowledge: pathology-specific, genomic-specific, and modality-common information from pathological WSIs and genomic profiles. Specifically, we leverage MLP layers as the modality-specific encoders $\mathrm{\Gamma}_{p}$ and $\mathrm{\Gamma}_{g}$ to extract specific representations $\mathrm{h}_{p}^{s}$ and $\mathrm{h}_{g}^{s}$, and a two-branch parallel MLP layer with a co-attention block as the modality-common encoder $\mathrm{\Gamma}_{c}$, which bridges the multimodal interactions and combines input features to generate two modality-common representations $\mathrm{h}_{p}^{c}$ and $\mathrm{h}_{g}^{c}$. The detailed operation of the modality-common encoder $\mathrm{\Gamma}_{c}$ can be formulated as:
\begin{equation}
    \mathrm{h}_{p}^{c}, \mathrm{h}_{g}^{c} 
    = \mathrm{\Gamma}_{c}\left( \mathrm{h}_{p}^{o}, \mathrm{h}_{g}^{o} \right)
    = \text{MLP}\left( \mathbb{A}^{\top} \right) \ast \mathrm{h}_{p}^{o},
      \text{MLP}\left( \mathbb{A} \right) \ast \mathrm{h}_{g}^{o},
\end{equation}
\begin{equation}
    \mathbb{A} = 
    \text{Linear} \left( \mathrm{h}_{p}^{o} \right)^{\top} 
    \cdot \text{Linear} \left( \mathrm{h}_{g}^{o} \right),
\end{equation}
where $\text{MLP}(\cdot)$ is composed of two fully-connected layers with layer normalization and GELU activation, $\mathbb{A}$ means the multimodal co-attention matrix, respectively. The $\ast$ is element-wise multiplication, while $\cdot$ denotes the matrix multiplication.

\subsection{Multimodal Feature Integration}

\subsubsection{Multimodal Representation Learning.} 
Inspired by previous work~\cite{songmultimodal}, we design a early-fusion based multimodal representation learning strategy to learn inter-modal interactions between histopathology and genomic profiles. Concretely, we first obtain the modality-specific representation $f_{s}^{o}$ and -common representation $f_{c}^{o}$ by concatenating their associated multimodal representations. Then, we develop a cross-attention module with two transformer decoders to integrate the modality-specific and -common representations into a joint embedding $f_{s}^{\prime}$ then mine their inner contextual information. This procedure can be formulated as:
\begin{equation}
    f_{s}^{\prime} = \text{CrossAttn} \left( f_{c}^{o}, f_{s}^{o}, f_{s}^{o} \right) 
    = \text{Softmax} \left( 
    \frac{\mathbf{W}_{q} \cdot f_{c}^{o} \cdot f_{s}^{o\top} \cdot \mathbf{W}_{k}^{\top}}{\sqrt{d_{k}}} \right) \mathbf{W}_{v} \cdot f_{s}^{o} 
\end{equation}
where $\mathbf{W}_{q}$, $\mathbf{W}_{k}$, $\mathbf{W}_{v} \in \mathbb{R}^{d_{k} \times d_{k}}$ are learnable weight matrices multiplied to the query $f_{c}^{o}$ and key-value pair $\left( f_{c}^{o},f_{c}^{o} \right)$, respectively. We then enhanced the resulting $f_{s}^{\prime}$ and $f_{c}^{\prime}$ by two transformer decoder $\mathcal{D}(\cdot)$ which can be formulated as: 
\begin{equation}
    \begin{aligned}
        & f_{s} = \mathcal{D}(f_{s}^{\prime}) 
        = \text{MSA}\Big( \text{LN} \left( f_{s}^{(2)} \right) \Big)+ f_{s}^{(2)}, \\
        & f_{s}^{(2)} = \text{PPEG} \left( f_{s}^{(1)} \right), 
        f_{s}^{(1)} = \text{MSA}\Big( \text{LN} \left( f_{s}^{\prime} \right) \Big) + f_{s}^{\prime},
    \end{aligned}
\end{equation}
where MSA and LN are Multi-head Self-Attention and Layer Norm, and we use the Nystrom attention~\cite{zhou2023cross} for MSA to mitigate heavy computation burden. $\text{PPEG}(\cdot)$ is the Pyramid Position Encoding Generator in~\cite{shao2021transmil} to capture contextual information. Likewise, we can obtain enhanced modality-common representation $f_{c}$ via $\mathcal{D}(\cdot)$.

\subsubsection{Deep Holistic Orthogonal Fusion.} 
We design a novel orthogonal fusion module to integrate modality-specific and -common representations, for modeling the holistic interactions across modalities. 
It takes $f_{s}$ and $f_{c}$ as inputs then calculates the projection $f_{s,proj}^{(i,j)}$ of each feature in specific representation onto the common representation $f_{c}$. 
Subsequently, we can compute the orthogonal component by:
\begin{equation}
    f_{s,orth}^{(i,j)} = f_{s}^{(i,j)} - f_{s,proj}^{(i,j)}, 
    f_{s,proj}^{(i,j)} = \frac{f_{s}^{(i,j)} \cdot f_{c}}{\left| f_{c} \right|^{2}} \cdot f_{c}
    = \frac{\sum_{k=1}^{K} f_{s,k}^{(i,j)} \cdot f_{c,k}}{\sum_{k=1}^{K} \left( f_{c,k} \right)^{2}} \cdot f_{c},
\end{equation}
where $\left| f_{c} \right|^{2}$ represents the $L_{2}$ norm of $f_{c}$. Here, we then leverage the pooling functionality to aggregate the concatenated tensor $f_{s,orth}^{(i,j)}$ as illustrated in the upper-right section of~\figurename~\ref{framework}. Lastly, a fully-connected layer is employed to produce the final multimodal representation $f_{mm}$ for survival outcome prediction.

\subsection{Training Objective}
The overall learning of the model is performed by minimizing:
\begin{equation}
   \mathcal{L}_{total} = \alpha \mathcal{L}_{sim} + \beta \mathcal{L}_{diff} + \gamma \mathcal{L}_{recon} + \mathcal{L}_{surv},
\end{equation}
Here, $\alpha$, $\beta$, and $\gamma$ are the positive hyper-parameters that determine the contribution of each regularization component to the overall training objective $\mathcal{L}_{total}$.
\subsubsection{Similarity Loss.} To reduce the discrepancy between the two modality-common representations and align them into the shared space, we introduce $L_{1}$ norm to measure the
discrepancy between the distribution of common representations.
\begin{equation}
   \mathcal{L}_{sim} = \mathcal{L}_{1} \left(\mathrm{h}_{p}^{o}, \mathrm{h}_{p}^{c} \right) = \frac{1}{n} ||\mathrm{h}_{p}^{o} - \mathrm{h}_{p}^{c} ||_1,
\end{equation}
\subsubsection{Difference Loss.} We further employ a difference loss to ensure that the
modality-common and -specific representations capture distinct aspects of the input. To this end, we introduce the Kullback-Leibler Divergence~\cite{kullback1951information} for this purpose.
\begin{equation}
   \mathcal{L}_{diff} = \mathcal{D}_{KL} \left(\mathrm{h}_{p}^{c}, \mathrm{h}_{p}^{s} \right) +                       \mathcal{D}_{KL} \left(\mathrm{h}_{g}^{c}, \mathrm{h}_{g}^{s} \right) = 
   \sum_{m}^{\left\{p,g\right\}} \left[ \mathrm{h}_m^c \log \left( \frac{\mathrm{h}_m^c}{\mathrm{h}_m^s} \right) \right],
\end{equation}
\subsubsection{Reconstruction Loss.} To prevent the learning of trivial representations, we first reconstruct the modality representations $\mathrm{h}_{p}^{o}$ and $\mathrm{h}_{g}^{o}$ using two MLP-based decoders to produce $\mathrm{h}_{p}^{r}$ and $\mathrm{h}_{g}^{r}$ two reconstructed representations. Then we apply mean squared error (MSE) loss as the modality reconstruction loss which ensures that the hidden representations capture the relevant details of each modality.
\begin{equation}
   \mathcal{L}_{recon} = 
   \frac{1}{2} \big( \mathcal{L}_{MSE}\left( \mathrm{h}_{p}^{o}, \mathrm{h}_{p}^{r} \right) + \mathcal{L}_{MSE}\left( \mathrm{h}_{g}^{o}, \mathrm{h}_{g}^{r} \right) \big)
   = \frac{1}{2} \sum_{m}^{\left\{p,g\right\}} \left( \mathrm{h}_m^o - \mathrm{h}_m^r \right)^2,
\end{equation}
\subsubsection{Survival Loss.} Following previous works~\cite{chen2021multimodal,liu2023mgct}, we utilize NLL (negative log-likelihood) survival loss~\cite{zadeh2020bias} as the $\mathcal{L}_{surv}$ for the survival prediction part. 

\section{Experiments and Results}
\subsection{Datasets} \label{sec:datasets}
Following~\cite{zhou2023cross}, we conducted extensive experiments on six cancer types from The Cancer Genome Atlas (TCGA) including Bladder Urothelial Carcinoma (BLCA), Breast Invasive Carcinoma (BRCA), Colon Adenocarcinoma \& Rectum Adenocarcinoma (COADREAD), Lung Adenocarcinoma (LUAD), Uterine Corpus Endometrial Carcinoma (UCEC), and Stomach Adenocarcinoma (STAD). We utilized RNA sequencing (RNA-seq), Copy Number Variation (CNV), and mutation status, and further categorized them into six genomic sub-sequences (protein kinases, tumor suppressor genes, oncogenes, cell differentiation markers, transcription, and cytokines and growth) as the genomic profile input of models.

\subsection{Implementation Details}
\subsubsection{Training settings.} 
We select a diverse set of baseline methods, including those focused on pathological WSIs, genomic data, and both modalities. The methods for comparison include: MLP, SNN~\cite{klambauer2017self}, Coxnnet~\cite{ching2018cox}, ABMIL~\cite{ilse2018attention}, DSMIL~\cite{li2021dual}, CLAM~\cite{lu2021data}, TransMIL~\cite{shao2021transmil}, DTFD-MIL~\cite{zhang2022dtfd}, M3IF~\cite{li2021multi}, MCAT~\cite{chen2021multimodal}, CMTA~\cite{zhou2023cross}, MoME~\cite{xiong2024mome}, MOTCat~\cite{xu2023multimodal}, SurvPath~\cite{jaume2024modeling}, and PORPOISE~\cite{chen2022pan}.
Following standard practice, we performed 5-fold monte-carlo cross-validation on each dataset to mitigate batch effects.
The mean Concordance index (C-index) with its standard deviation (std) are reported for a comprehensive performance comparison. Moreover, Kaplan-Meier~\cite{bland1998survival} and T-test analyses are employed to assess the significance of differences in survival prediction between high- and low-risk groups.

\subsubsection{Hyper-parameters.}
We trained all models with the Adam~\cite{kingma2014adam} optimizer for a total of 20 epochs. The learning rate and weight decay were set to $2\times10^{-4}$ and $1\times10^{-5}$, respectively. 
During training of SurvPath~\cite{jaume2024modeling}, we randomly sample 4,096 patches per WSI to avoid memory overflow issues.
We set the hyper-parameters $\alpha$ and $\beta$ as $1\times10^{-4}$, $\gamma$ as 1 for our training loss function in all cancer datasets.

\newcommand{\myboxsize}{1\textwidth}
\newcommand{\myarraystretch}{1.3}
\begin{table*}[t]
    \center
    \caption{The performance of different approaches on six public TCGA datasets. ``P.'' indicates whether to use pathological images and ``G.'' indicates whether to use genomic profiles. The best and second best results are highlighted in \textbf{bold} and \underline{underlined}.}
	\renewcommand{\arraystretch}{\myarraystretch}
	\setlength\tabcolsep{6pt}
	\centering
	\resizebox{\myboxsize}{!}
    {
    \begin{tabular}{lcccccccc}
        \toprule[1pt]
        {\multirow{2}{*}{Methods}} & \multicolumn{2}{c}{Modality} & BLCA & BRCA & COADREAD & LUAD & UCEC & STAD 
        \\ \cmidrule(lr){2-3} 
        & P. & G. & ($n=373$) & ($n=956$) & ($n=340$) & ($n=453$) & ($n=480$) & ($n=349$) \\ 
        \midrule
        MLP &  & $\checkmark$ 
        & 0.611 $\pm$ 0.030 & 0.619 $\pm$ 0.053 & 0.675 $\pm$ 0.049 
        & 0.619 $\pm$ 0.044 & 0.675 $\pm$ 0.069 & 0.607 $\pm$ 0.045 \\
        SNN~\cite{klambauer2017self}
        &  & $\checkmark$ 
        & 0.625 $\pm$ 0.054 & 0.621 $\pm$ 0.051 & 0.606 $\pm$ 0.087 
        & 0.615 $\pm$ 0.044 & 0.703 $\pm$ 0.072 & 0.603 $\pm$ 0.052 \\ 
        Coxnnet~\cite{ching2018cox}
        &  & $\checkmark$ 
        & 0.591 $\pm$ 0.061 & 0.593 $\pm$ 0.067 & 0.601 $\pm$ 0.059 
        & 0.605 $\pm$ 0.044 & 0.535 $\pm$ 0.051 & 0.523 $\pm$ 0.055 \\
        \midrule
        ABMIL~\cite{ilse2018attention}
        & $\checkmark$ & 
        & 0.640 $\pm$ 0.032 & 0.642 $\pm$ 0.030 & 0.703 $\pm$ 0.064
        & 0.607 $\pm$ 0.041 & 0.701 $\pm$ 0.038 
        & \underline{0.638 $\pm$ 0.028} \\
        DSMIL~\cite{li2021dual}
        & $\checkmark$ & 
        & 0.636 $\pm$ 0.031 & 0.657 $\pm$ 0.037 & 0.684 $\pm$ 0.071
        & 0.591 $\pm$ 0.026 & 0.729 $\pm$ 0.063 & 0.630 $\pm$ 0.032 \\
        CLAM~\cite{lu2021data}        
        & $\checkmark$ & 
        & 0.618 $\pm$ 0.042 & 0.593 $\pm$ 0.077 & 0.643 $\pm$ 0.108
        & 0.580 $\pm$ 0.029 & 0.647 $\pm$ 0.037 & 0.634 $\pm$ 0.054 \\
        TransMIL~\cite{shao2021transmil}
        & $\checkmark$ & 
        & 0.640 $\pm$ 0.039 & 0.590 $\pm$ 0.058 & 0.684 $\pm$ 0.024
        & 0.592 $\pm$ 0.048 & 0.705 $\pm$ 0.027 & 0.604 $\pm$ 0.054 \\
        DTFD-MIL~\cite{zhang2022dtfd}
        & $\checkmark$ & 
        & 0.610 $\pm$ 0.025 & 0.625 $\pm$ 0.060 & 0.633 $\pm$ 0.061 
        & 0.590 $\pm$ 0.033 & 0.663 $\pm$ 0.016 & 0.580 $\pm$ 0.033 \\
        \midrule 
        M3IF~\cite{li2021multi}
        & $\checkmark$ & $\checkmark$ 
        & 0.662 $\pm$ 0.043 & 0.639 $\pm$ 0.062 
        & \underline{0.710 $\pm$ 0.043}
        & 0.628 $\pm$ 0.050 & 0.724 $\pm$ 0.103 & 0.607 $\pm$ 0.044 \\
        MCAT~\cite{chen2021multimodal}
        & $\checkmark$ & $\checkmark$  
        & 0.677 $\pm$ 0.062 
        & \underline{0.691 $\pm$ 0.041} & 0.649 $\pm$ 0.052 
        & 0.675 $\pm$ 0.039 & 0.658 $\pm$ 0.055 & 0.586 $\pm$ 0.027 \\
        CMTA~\cite{zhou2023cross}
        & $\checkmark$ & $\checkmark$  
        & 0.683 $\pm$ 0.016 & 0.667 $\pm$ 0.033 & 0.678 $\pm$ 0.040
        & 0.648 $\pm$ 0.036 & 0.740 $\pm$ 0.066 & 0.584 $\pm$ 0.023 \\
        MoME~\cite{xiong2024mome}
        & $\checkmark$ & $\checkmark$ 
        & 0.650 $\pm$ 0.033 & 0.659 $\pm$ 0.032 & 0.692 $\pm$ 0.038 
        & 0.650 $\pm$ 0.024 & 0.705 $\pm$ 0.041 & 0.630 $\pm$ 0.039 \\
        MOTCat~\cite{xu2023multimodal}
        & $\checkmark$ & $\checkmark$ 
        & \underline{0.683 $\pm$ 0.025}
        & 0.675 $\pm$ 0.021 & 0.618 $\pm$ 0.030 
        & 0.667 $\pm$ 0.027 & 0.685 $\pm$ 0.052 & 0.596 $\pm$ 0.028\\
        SurvPath~\cite{jaume2024modeling}  
        & $\checkmark$ & $\checkmark$  
        & 0.673 $\pm$ 0.011 & 0.685 $\pm$ 0.024 & 0.650 $\pm$ 0.024
        & \underline{0.676 $\pm$ 0.036}
        & 0.737 $\pm$ 0.049 & 0.622 $\pm$ 0.045 \\
        PORPOISE~\cite{chen2022pan}   
        & $\checkmark$ & $\checkmark$  
        & 0.651 $\pm$ 0.046 & 0.624 $\pm$ 0.048 & 0.702 $\pm$ 0.045 
        & 0.642 $\pm$ 0.042 
        & \underline{0.737 $\pm$ 0.097} & 0.612 $\pm$ 0.038 \\
        \midrule
        \textbf{MurreNet}    
        & $\checkmark$  & $\checkmark$ 
        & \textbf{0.710 $\pm$ 0.030}
        & \textbf{0.718 $\pm$ 0.016}
        & \textbf{0.725 $\pm$ 0.044}
        & \textbf{0.691 $\pm$ 0.040}
        & \textbf{0.752 $\pm$ 0.075}
        & \textbf{0.651 $\pm$ 0.062}
        \\
        \bottomrule[1pt]
    \end{tabular}
    }\label{tab:comparison}
\end{table*}

\subsection{Comparison with State-of-the-art Methods}
We conducted extensive experiments compared with 15 bleeding-edge survival prediction methods using identical settings on six TCGA datasets. As shown in \tablename~\ref{tab:comparison}, MurreNet achieves superior performance on all cancer types. Notably, most multimodal methods steadily outperform unimodal methods, highlighting the crucial contribution of integrating information from both modalities. Against MCAT~\cite{chen2021multimodal}, the current SOTA multimodal method, our model achieves the performance increases of 4.87\% on BLCA, 3.91\% on BRCA, 11.71\% on COADREAD, 2.37\% on LUAD, 14.29\% on UCED, and 11.09\% on STAD, respectively. This suggests that survival analysis using multimodal data should focus on the holistic interactions between pathology and genomic information, rather than merely exploiting the abundant shared correlations. Additionally, our model also outperforms CMTA~\cite{zhou2023cross}, further emphasizing the importance of disentangling and reintegrating modality-specific and -common knowledge.
Finally, our model consistently outperforms other SOTA multimodal methods by a large margin, including M3IF~\cite{li2021multi}, MoME~\cite{xiong2024mome}, MOTCat~\cite{xu2023multimodal}, SurvPath~\cite{jaume2024modeling} and PORPOISE~\cite{chen2022pan}.

\subsection{Ablation Study}
To thoroughly validate the impact of each proposed component in our model, we perform an ablation study on six TCGA datasets by breaking down our model into several submodels. 
We begin with a multimodal survival baseline A, which simply concatenating pathology and genomic features for survival analysis. 
Next, we add the MRD module into model A as model B, followed by the incorporation of DHOF strategy as model C, and the addition of $\mathcal{L}_{sim}$ (D), $\mathcal{L}_{diff}$ (E), and $\mathcal{L}_{recon}$ training loss as our final model F.
Quantitative results are presented in \tablename~\ref{tab:ablation}.
By comparing models A and B, we observe that our MRD design significantly enhances modality-specific and -common representation representation learning, leading to performance improvements of 1.71\% on BLCA, 2.07\% on LUAD, 2.38\% on UCEC, and 1.34\% on STAD, respectively.
The DHOF strategy in model C further improves the results, contributing to more accurate survival predictions.
Additionally, the inclusion of the training strategies $\mathcal{L}_{sim}$, $\mathcal{L}_{diff}$, and $\mathcal{L}_{recon}$ continues to boost performance, as indicated by the increase in C-index from model D to F.
This suggests that the learning constraints on modality similarity, difference, and reconstruction play a crucial role in facilitating holistic multimodal interaction modeling for cancer prognosis.

\begin{table*}[t]
	\centering
	\caption{Quantitative results for ablation study on four TCGA cancer datasets.}\label{tab:ablation}
	\renewcommand{\arraystretch}{\myarraystretch}
	\setlength\tabcolsep{6pt}
	\resizebox{\myboxsize}{!}
	{\begin{tabular}{cccccccccc}
    \toprule[1pt]
    \multirow{2}{*}{Model} & \multicolumn{5}{c}{{Designs in our model}} & \multicolumn{4}{c}{Datasets} \\
    \cmidrule(lr){2-6}
    \cmidrule(lr){7-10}
    & MRD & DHOF & $\mathcal{L}_{sim}$ & $\mathcal{L}_{diff}$ & $\mathcal{L}_{recon}$ 
    & BLCA & LUAD & UCEC & STAD \\ \midrule
    A & & & & & 
    & 0.645 $\pm$ 0.017 & 0.627 $\pm$ 0.053 & 0.673 $\pm$ 0.047 & 0.598 $\pm$ 0.026 \\
    B & \checkmark &  & & & 
    & 0.656 $\pm$ 0.031 & 0.640 $\pm$ 0.011 & 0.689 $\pm$ 0.010 & 0.606 $\pm$ 0.042 \\
    C & \checkmark & \checkmark & & & 
    & 0.673 $\pm$ 0.024 & 0.652 $\pm$ 0.041 & 0.701 $\pm$ 0.052 & 0.617 $\pm$ 0.034 \\
    D & \checkmark & \checkmark & \checkmark & & 
    & 0.695 $\pm$ 0.028 & 0.674 $\pm$ 0.039 & 0.733 $\pm$ 0.060 & 0.636 $\pm$ 0.038 \\
    E & \checkmark & \checkmark & \checkmark & \checkmark & 
    & 0.701 $\pm$ 0.030 & 0.680 $\pm$ 0.042 & 0.746 $\pm$ 0.063 & 0.644 $\pm$ 0.040 \\
    F & \checkmark & \checkmark & \checkmark & \checkmark & \checkmark 
    & \textbf{0.710 $\pm$ 0.030} & \textbf{0.691 $\pm$ 0.040} 
    & \textbf{0.752 $\pm$ 0.075} & \textbf{0.651 $\pm$ 0.062} \\
	\toprule[1pt]
	\end{tabular}}
\end{table*}

\subsection{Risk Stratification}
\begin{figure}[h]
    \center
    \centering
    \includegraphics[width=0.99\textwidth]{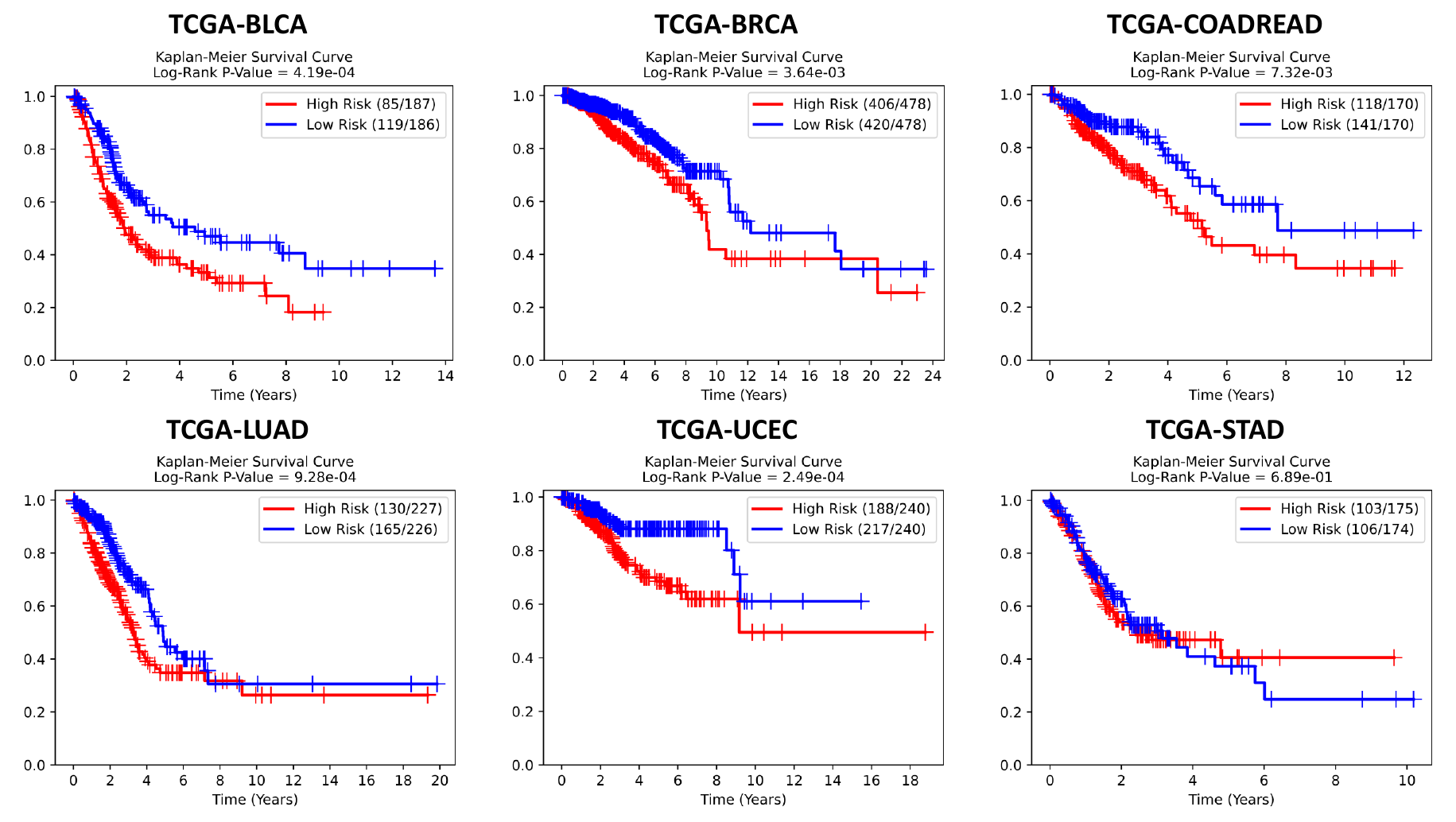}
    \caption{Kaplan-Meier survival curves of the proposed model on six cancer datasets.}
    \label{fig:kmcurve}
\end{figure}

To further substantiate the efficacy of our model for survival analysis, we stratify all patients into low- and high-risk groups on the median of the predicted risk scores generated by our model across six TCGA cohorts. Subsequently, we utilize Kaplan-Meier~\cite{bland1998survival} survival curve to visually depict the survival events of all the patients. To evaluate the statistical significance of the differences between these groups, we apply the Log-rank test~\cite{bland2004logrank}, where a p-value of 0.05 or below is deemed statistically significant.  As shown in \figurename~\ref{fig:kmcurve}, the p-values for our model across six cohorts are consistently well below 0.05, underscoring the robustness of MurreNet in effective risk stratification and reinforcing its clinical potential.

\section{Conclusion}
In this paper, we propose the Multimodal Representation Decoupling framework (MurreNet) to integrate genomics and pathological WSIs for survival analysis. We innovatively propose MRD module which factorizes modalities into modality-common and modality-specific information to reduce the modality redundancy. To enhance learning constraints, we incorporate a combination of losses for modality similarity, difference, reconstruction and survival prediction task. Finally, we design the DHOF strategy to enable multimodal feature fusion and model the intricate inter- and intra-modality correlations.
Extensive experiments on six TCGA datasets indicate effectiveness of our proposed MurreNet.

\subsubsection{Acknowledgments.} 
This work is supported by the National Key R$\&$D Program of China (No.2023YFC3402800), National Natural Science Foundation of China (Nos.82441029, 62171230, 62101365, 92159301, 62301263, 62301265, 62302228, 82302291, 82302352, 62401272), Jiangsu Provincial Department of Science and Technology's major project on frontier-leading basic research in technology (No.BK2023200).

\subsubsection{Disclosure of Interests.} 
The authors have no competing interests to declare that are relevant to the content of this article.

%
%

\bibliographystyle{splncs04}
\bibliography{Paper-0057}
\end{document}